\documentclass[aps,prl]{revtex4}
\usepackage{amsfonts}
\newcommand{\cA}{{\mathcal A}}
\newcommand{\fA}{{\mathfrak A}}
\newcommand{\cB}{{\mathcal B}}

\newcommand{\cF}{{\mathcal F}}
\newcommand{\cH}{{\mathcal H}}

\newcommand{\bR}{{\mathbb{R}}}
\newcommand{\bC}{{\mathbb{C}}}

\newcommand{\bZ}{{\mathbb{Z}}}

\newcommand{\jed}{{\mathbb{I}}}
\begin{document}

\setlength{\baselineskip}{2\baselineskip}

\begin{center}
\vspace*{15mm}
{\large On non-completely positive quantum dynamical maps on spin chains.\\}

\vspace{2cm}

\textsc{{{ W. A. Majewski}}}\\

\vspace{2cm}
\textsc{{}\\
Institute of Theoretical Physics and Astrophysics\\
Gda{\'n}sk University\\
Wita Stwosza~57\\
80-952 Gda{\'n}sk, Poland}\\
\textit{E-mail address:} \texttt{fizwam@univ.gda.pl}\\

\vspace*{3cm}

Abstract:
The new arguments based on the Majorana fermions model indicating that non-completely positive maps can describe open quantum evolution are presented.

 \end{center}

\vskip 3cm

\newpage

\section{Introduction}
The aim of this note is to provide the contribution to the long debate on the nature of quantum maps, see \cite{S1}-\cite{M22}. At the heart of this debate
is a hypothesis, supported by some arguments, that only a subclass of linear unital positive maps - completely positive maps ({\it CP maps}) - is relevant for a description of any physical evolution. This claim is 
challenged by several authors, see \cite{P15}-\cite{B?} and \cite{S17}-\cite{M22}. Here, we wish to present the model of quantum evolution on a lattice supporting the claim that completely positive maps do not exhaust all possible dynamical behavior that an open quantum dynamical system may exhibit. To be more precise, let us recall that the main approach to the description of time evolution of open systems is based on the so called Nakajima-Zwanzig projection technique (see \cite{Nak}-\cite{Rau}). The main idea of this approach says that if we are interested in a subsystem defined by some properly chosen family of observables, the dynamics of the full system should be projected onto the considered subsystem, so in other words we neglect some correlations and non relevant parameters (usually attributed to reservoir). Subsequently, an application of certain approximations leads to dynamical semigroups describing the time evolution of the relevant subsystem. {\it Thus, elimination of unmonitored degrees of freedom in the space of observables carried out within the scheme of the projection technique yields the dynamics of the open systems}.

Our first remark is that such a procedure, in general, can be irreversible in the sense that starting  from the reduced dynamics one can not always expect the possibility to reconstruct the full dynamics. In more mathematical terms, it means that not every reduced dynamical map is extensible to the dynamics of the full system. It should be  stressed that any completely positive (CP for short) map is extensible (see  \cite{Av}; for a one parameter CP semigroup see \cite{EvaL}).

The next remark is that individual properties of any system are encoded in the proper choice of the family of its observables. Consequently, each system (or any subsystem) should have its observables in an algebra with specific properties. A quick review of axioms of Quantum Mechanics (see \cite{Dirac}-\cite{CMP}) indicates that $C^*$-algebraic structure for the set of observables is taken for its mathematical convenience only.
We recall that $C^*$-algebra is a generalization of the algebra of linear bounded operators on a Hilbert space. The properties of operator addition, multiplication, hermitian conjugation, and the norm are axiomatized in a prescribed way.
 On the other hand, it seems that the structure of Jordan algebra is a more fundamental concept. 
 Here, a Jordan algebra $\frak{B}$ (to be more precise we restrict ourselves to JC-algebras) stands for a norm closed real vector space of bounded self-adjoint operators on a complex Hilbert space which is also closed with respect to anticommutator, i.e. under the Jordan product, i.e.
 \begin{equation}
 a \circ b = \frac{1}{2}(ab + ba), \quad a,b \in \frak{B}
 \end{equation}
 
However, in all physical models, the Jordan algebra of a physical system can be embedded in the corresponding $C^*$-algebra structure. We will use this remark in the construction of our model.

These observations can be combined with the very deep St{\o}rmer's result stating that the nature of positivity of a certain class of linear maps on an algebra can be linked with algebraic properties of their images \cite{St}. 
To quote this result we recall that the set of all positive maps, in general, contains a {\it proper} subset of maps which are called decomposable. These maps 
 $\alpha: \cA \to \cB(\cH)$,
can be represented as a sum of CP map and a co-CP map (co-CP map is the composition of CP map with the transposition $\tau$), i.e. $\alpha = \alpha_1 + \tau \cdot \alpha_2$ where $\alpha_i$ are CP maps.
St{\o}rmer has proved that a linear unital 
positive map $P$ on the algebra $\cA$ such that $P \cdot P = P$ (a projection) is a decomposable map if only if the Jordan algebra associated with the image of the map $P(\cA)$ is the reversible one (so it is closed with respect to the product $\{a_1,a_2,...,a_n\} \equiv a_1a_2...a_n + a_n...a_2a_1$, $n=1,2,..$). 
In other words, if $P(\cA)$ is not a reversible Jordan algebra then the projection $P$ is not a decomposable map.
This observation leads to a natural question about examples of non-reversible Jordan algebras. To answer this question, we recall the construction of spin factors. To this end, we define a spin system $\frak{S}$ to be a collection of nontrivial symmetries, i.e. anticommuting operators on a Hilbert space such that 
\begin{equation}
s \cdot s = \jed, \quad s=s^*, \quad s\neq \pm \jed, \quad s\circ t =0
\end{equation}
for any $s,t \in \frak{S}$. Clearly, on two dimensional Hilbert space $\bC^2$, Pauli matrices provide an example of spin system.
Then, the spin factor $\frak{F}$ is defined as the smallest unital JC algebra containing the spin system.

Most of spin factors are not reversible. More precisely, only spin factors of dimension less than 4 are reversible. The six dimensional spin factor admits both reversible and non-reversible representations. The rest are non-reversible. Summarizing: if one constructs a projection onto a spin factor of dimension larger than six, such projector is a positive, {\it nondecomposable} map.
We will use this observation in Section 2.

Turning to the question of dynamical maps on a subsystem we emphasize that exactly the concept of a projection map is in the heart of  Nakajima-Zwanzig's prescription (briefly reviewed above) and the Nakajima-Zwanzig approach is the basic ingredient of the theory of open systems! Consequently, properties of reduced dynamical maps may depend on the corresponding properties of  projections.

Taking into account these remarks in the description of wide class of lattice models we will show, in the next section, that there is a possibility to present concrete models of physical systems with the reduced dynamics given by linear positive unital non-decomposable maps. The paper is closed with concluding remarks given in the third section.

\section{Model}

Let $\Sigma$ stand for the one dimensional lattice system. Assume that at each site $x$ of the lattice $\Sigma$ the algebra of observables $\fA_x$ is given by the set of all $2 \times 2$ matrices, i.e. $\fA_x \equiv M_2(\bC)$. The algebra of observables $\fA_{\Lambda}$ associated with some bounded region $\Lambda$ of the lattice $\Sigma$ is given by $\fA_{\Lambda} = \otimes_{x \in \Lambda} \fA_x \equiv \otimes_{x \in \Lambda}M_2(\bC)$. Let $\Lambda \subseteq \Lambda^{\prime} \subset \bZ$, then $\fA_{\Lambda}$ can be embedded into $\fA_{\Lambda^{\prime}}$ by $I_{\Lambda,\Lambda^{\prime}} : \fA_{\Lambda} \to \fA_{\Lambda} \otimes \jed_{\Lambda^{\prime} \setminus \Lambda}$. Finally, the algebra of the whole system $\Sigma$ is given by the inductive limit $\fA = \lim_{\Lambda \to \bZ} \left(\fA_{\Lambda},I_{\Lambda,\Lambda^{\prime}}\right)$.

To describe systems with interactions we need to introduce a notion of
an interaction potential. To this end we denote by $\cF$ the family of bounded regions in $\bZ$. 
A family $\Phi \equiv \{ \Phi_X\in\fA_X\}_{X\in \cF}$ of selfadjoint operators 
such that
\begin{equation}
\Vert \Phi \Vert_1 \equiv \sup_{i\in\bZ} \sum_{X\in\cF\atop{X\ni i}}
 \Vert \Phi_X\Vert < \infty
\end{equation}
will be called a (Gibbsian) potential. 
A potential $\Phi \equiv \{ \Phi_X \}_{X\in\cF}$ is
of {\it finite range} $R\ge 0$, iff 
$ \Phi_X  =0$ for all $X\in \cF,\quad \mathrm{diam}(X) >R$.
The corresponding Hamiltonian $H_\Lambda$, $\Lambda\in\cF$, is defined by
\begin{equation}
H_\Lambda \equiv H_\Lambda (\Phi )\equiv \sum_{X\subset\Lambda} \Phi_X
\end{equation}

In particular, it is an easy observation that any finite range and  nearest-neighbour interactions fall into the considered class 
of systems. The Hamiltonian dynamics for a finite region $\Lambda$ is defined  as the following automorphism group 
associated to potential $\Phi $
\begin{equation}
\alpha_t^\Lambda(a) \equiv e^{+it H_\Lambda} a e^{-it H_\Lambda}
\end{equation}
for $a \in \fA$.
If the potential $\Phi \equiv \{ \Phi _X\}_{X\in\cF}$ satisfies
also
\begin{equation}
\label{war1}
||\Phi||_{exp }
\equiv 
\sup_{i\in\bZ}\sum_{X\in\cF\atop{X\ni i}} e^{\lambda|X|}||\Phi_X||
<\infty
\end{equation}
for some $\lambda >0$, then the following limit exists,  \cite{BR}, 
\begin{equation}
\alpha_t (a) \equiv \lim  \alpha_t^\Lambda (a)
\end{equation}
for every $a\in\fA_0 \equiv \cup_{\Lambda \in \cF} \fA_{\Lambda}$.
Consequently, the specification of local interactions leads
to the well defined global dynamics provided that (\ref{war1}) is valid. Hence, we defined the dynamical system associated to $\Sigma$
\begin{equation}
\label{1}
\left(\fA, \alpha_t, t \in \bR \right)
\end{equation}

Obviously, as each $\alpha_t, t \in \bR$ is a $^*$-automorphism, the dynamics in (\ref{1}) is given by completely positive maps.

 Now, we wish to define a fermionic subsystem of $\Sigma$ associated to a fixed arbitrary bounded region $\Lambda_0 \equiv \{0,..., +n\}$ of $\bZ$, i.e. a set of anticommuting  observables associated with the sites $\{0,1,2,...,(n-1),n\}$ will be singled out. Note, that spins at different sites commute. In the Fermion algebra, the observables at different sites anticommute instead of commuting as in the algebra of spins. This is achieved by applying the idea of the Jordan-Wigner transformation \cite{JW}. To this end,
firstly, we recall definitions of Pauli matrices:

$$\sigma_0=\jed,\quad\sigma_1=\left[\begin{array}{rr}1&0\\0&-1\end{array}\right],\quad
\sigma_2=\left[\begin{array}{rr}0&1\\1&0\end{array}\right],\quad
\sigma_3=\left[\begin{array}{rr}0&i\\-i&0\end{array}\right].$$
They form a spin system i.e.
\begin{equation}
\sigma_i\circ\sigma_j=\delta_{ij}\jed,\quad i,j=1,2,3.
\end{equation}
where $\circ$ stands for the Jordan product.
Further, introducing
\begin{equation}
\sigma^x_{\pm} = \frac{1}{2}\left(\sigma^x_1 \pm i \sigma_2^x\right)
\end{equation}
where $x \in \Lambda_0$ and $\sigma_i^x$ denotes the Pauli $\sigma_i$ matrix in the $x$-site, one defines
\begin{equation}
a_x= \sigma_3^{0} \otimes \sigma_3^{1} \otimes ...\otimes\sigma_3^{x-1}\otimes\sigma_-^x, \quad
a_x^*= \sigma_3^{0} \otimes \sigma_3^{1} \otimes ...\otimes\sigma_3^{x-1}\otimes\sigma_+^x
\end{equation}
Frequently, these operators are used to be written as
\begin{equation}
a_x = \sigma_-^x \cdot\Pi_{y=0}^{x-1}\sigma_3^y, \qquad a_x^*= \sigma_+^x \cdot\Pi_{y=0}^{x-1}\sigma_3^y 
\end{equation}
Clearly
\begin{equation}
\{a_x,a_y^* \} = \delta_{xy}, \quad \{\sigma_+^x, \sigma_-^x \} = \jed, \quad [\sigma_+^x, \sigma_-^x] = - \sigma_3^x
\end{equation}
so $a_x, a^*_y$ satisfy canonical anticommutation relations.  

To legitimize our construction that there is a spin factor describing a set of observables we
note, following quantum field (or harmonic oscillator) ideas, that one can interpret
\begin{equation}
\phi(f) \equiv \frac{1}{\sqrt{2}}\left(a(f) + a^*(f)\right), \quad \pi(f) \equiv \frac{1}{\sqrt{2}i}\left(a(f) - a^*(f)\right)
\end{equation}
as ``coordinate'' and ``momentum'' observables where $f$, in the context of quantum field theory, is a one particle wave function. In the abstract description of canonical anticommutation algebra (see \cite{BR}, \cite{Emch}), such $f$ is considered as an element of an index set (usually associated with some localization). Here, we adopt this way of thinking. 
Thus, it is natural to interpret the following set of operators
\begin{equation}
\phi \to \frac{1}{\sqrt{2}}\left(\sigma_+^x + \sigma_-^x\right)\cdot \Pi_{y=0}^{x-1} \sigma_3^y =
\frac{1}{\sqrt{2}}\sigma_1^x \cdot \Pi_{y=0}^{x-1} \sigma_3^y
\end{equation}
and
\begin{equation}
\pi \to \frac{1}{\sqrt{2}i}\left(\sigma_+^x - \sigma_-^x\right)\cdot \Pi_{y=0}^{x-1} \sigma_3^y =
\frac{1}{\sqrt{2}}\sigma_2^x \cdot \Pi_{y=0}^{x-1} \sigma_3^y
\end{equation}
where $x \in \Lambda_0$,
as representatives of fermionic observables at $x$, $ x \in \Lambda_0$. 
Now, it is easy to observe  that $\{ \sigma_1^x \cdot \Pi_{y=0}^{x-1} \sigma_3^y, \quad \sigma_2^x \cdot \Pi_{y=0}^{x-1} \sigma_3^y \}_{x \in \Lambda_0}$ form a spin system, thus they generate a spin factor (see \cite{HOS}).

Consequently, we singled out the set of observables
forming the subsystem $\Sigma_0$ associated with $\Lambda_0$. It should be noted that such approach to fermionic fields goes back at least to I. E. Segal \cite{Bon}, \cite{BSZ}. On the other hand, such operators called {\it Majorana fermions}  play the important role in the fermionic quantum computation, see \cite{Kit}, \cite{Vin}, e.g. a decoherence-free quantum memory can be described by certain systems of Majorana fermions (\cite{Kit}). It seems that such memory could be a key for the theory of quantum computer.

But, as we noted,  the Jordan algebra (not $C^*$-algebra!) generated by this set of selfadjoint operators is the non-reversible Jordan algebra (for $n$ larger than 2) -  it is a spin factor. Let us denote this Jordan algebra by $\fA_0$ and define the projector $P: {\fA}_{s.a.} \to \fA_0$, i.e. the linear unital positive map such that $P\cdot P = P$. It is important to note that such the projection exists by the Effros-St{\o}rmer result \cite{ESt} and it is a positive, unital {\it non-decomposable} map by another St{\o}rmer's result \cite{St}. Moreover, the projector $P$ does not depend on any reference state (separability of the reference state is the crucial assumption to get a CP reduced dynamical map, \cite{P15}, \cite{S17}).

Finally, let us turn to the reduced time evolution of observables relevant to subsystem $\Sigma_0$. The reduced (according to the projection technique) dynamical maps $\alpha^{\Sigma_0}_t : \fA_0 \to \fA_0$ are of the form
\begin{equation}
\alpha^{\Sigma_0}_t = P \cdot \alpha_t, \qquad t \in \bR^+
\end{equation}
It is obvious, that for each $t \in \bR^+$ {\it the maps $\alpha^{\Sigma_0}_t$ are positive unital and non-decomposable.}
We close this section with a remark that the presented model has many straightforward generalizations, e.g. to $n$-dimensional lattice models with $n\geq 1$.

\section{Conclusions}

 In general, neglecting some degrees of freedom and/or
leaving out certain correlations can lead to ``non-reversible'' procedure which spoils the extensibility of dynamics
of the selected observables.
It is worth pointing out that only CP maps always possess the proper extensions (see \cite{Av}, \cite{EvaL}; and such extensibility is the main ingredient of the so called  Lindblad argument in favour of CP maps \cite{L5}).
If the monitored observables could be embedded into a nuclear $C^*$-algebra, thus inclusion of a certain number of non-physical variables would be admissible, then the resulting dynamical maps are extensible \cite{Ste}. However, in general, such extensions are far from being the Hamiltonian type evolution.
The question of extension of dynamics is not the only one problem of extensibility which can be posed.
For example, describing fermionic models one can single out two sets of observables (Jordan algebras), $\fA_{I_1}$ and $\fA_{I_2}$, associated with two disjoint sets of sites $I_1 = \{ n_1,...n_k \}$ and $I_2 = \{m_1, ..., m_l \}$. However, a joint extension of states of $\fA_{I_1}$ and $\fA_{I_2}$ does not need, in general, to exist (see \cite{Ar1}, \cite{Ar2}).
Hence, we conclude that extensibility of both dynamics and states is {\it the very subtle and important question}. 
On the other hand, not taking into account this question can lead to serious difficulties as for example in \cite{SPLett}. We note that arguments based on the assignment map (see \cite{A20}, and \cite{SPLett}) can not be applied to the considered lattice model as the concept of this map is based on the assumption that a dynamical map of subsystem can be always be traced back to the full (hamiltonian) evolution.

Secondly, we want to point out that the projection $P$ considered in the second Section is not, in general, equal to the Umegaki's conditional expectation (see \cite{Ume}, and \cite{Tak}).
Namely, by definition, the image of a conditional expectation is a $C^*$-algebra. 
$P$ is also not equal to the generalized conditional expectation in the sense of Accardi-Cecchini \cite{AC}.
Both types of conditional expectations would 
guarantee the complete positivity of such the reduced dynamical maps. So, if one would pay the price and enlarge the set of selected observables with certain amount of non-physical quantities then the reduced dynamics to such enlarged subsystem would be completely positive. Therefore, we conclude that {\it the claim of complete positivity of any dynamical map is based on its mathematical convenience but, on purely physical grounds, CP-property does not seem to be absolutely necessary.}
Moreover, there are some concrete models, e. g. a three-qubit system, showing evidences in favour of non-CP parts of quantum maps (see \cite{Wein}).
 
Thirdly, an analysis of positive maps tempts to relate the obtained results to entanglement. In the standard setting, CP maps (decomposable, positive) correspond, by duality, to the set of all states (PPT, separable states respectively) and this correspondence is related to the tensor product structure. To indicate, without details, that one can expect similar ``tensor product structure'' in the considered model firstly we note that the the tensor product of JC-algebras is defined in terms of the corresponding (universal enveloping) $C^*$-algebras. Secondly, the presented model is defined in terms of {\it one-sided} finite lattices. The important point to note is that for a one sided lattice, {\it contrary to the case of two sided lattice}, the identifications of spin (Pauli) algebra $\otimes_{i=0}^n M_2(\bC)$ ($\otimes_{i=0}^{n+1} M_2(\bC)$)
with the fermion algebra $\cA^F_{[0,n]}$ generated by $\{a_i, a^*_j \}_{i,j=0}^n$ ($\cA^F_{[0,n+1]}$ respectively) are compatible with the embeddings  $\cA^F_{[0,n]}\subset \cA^F_{[0,n+1]}$ and $\otimes_{i=0}^n M_2(\bC) \subset \otimes_{i=0}^{n +1} M_2(\bC)$. Consequently, the (quasi)local structures of spin Pauli algebras and Majorana fermions algebras do not differ much. But to speak about entanglement one should split a system into two separate parts. However, this procedure, for Majorana fermions, leads to some problems. Namely, operators $a_x$, $a^*_x$ depend on all sites $x=0,1,...,x-1$. This can be taken as an argument that only ``even'' elements are relevant. 
Furthermore, turning to states on fermions, due to the extension problems mentioned in the first paragraphs
of this section, the concept of entanglement (separability) demands some elaboration (cf \cite{Moriya}).
Summarizing, in the restricted form, one can speak about entanglement for the considered model.
 
Finally, we stress that in Section 2 we got the one parameter family of quantum non-decomposable dynamical maps. To go further and to obtain a dynamical semigroup in the rigorous way, it would be necessary to perform other steps, for example to leave out the ``memory'' terms. But, as far as we know, this question is at the present far from being solved. However, for many physical problems, models with a discrete time evolution are acceptable.

\vskip 0.3cm
{\bf Acknowledgements:} 
The author thanks Y. S. Weinstein for his remarks. I also thank the anonymous referees for their comments that greatly improved the presentation of the paper.
This work was partially supported by KBN grant PB/1490/PO3/2003/25 and SCALA (IST-2004-015714)


\begin{thebibliography}{99}
\bibitem{S1}E. C. G. Sudarshan, P. M. Mathews, J. Rau, Phys. Rev. 121 (1961) 920.

\bibitem{K9} K. Kraus, Ann. Phys. 64 (1971) 311.

\bibitem{K4} A. Kossakowski, Rep. Math. Phys. 3 (1972) 119.

\bibitem{L5} G. Lindblad, Commun. Math. Phys. 48 (1976) 119.

\bibitem{D2} E. B. Davies, {\it Quantum Theory of Open systems}, Academic Press, New York, 1976.

\bibitem{G6} V. Gorini, A. Kossakowski, E. C. G. Sudarshan, Rep. Math. Phys. 13 (1978) 149.

\bibitem{S3} H. Spohn, Rev. Mod. Phys. 52 (1980) 569

\bibitem{K10} K. Kraus, {\it States, Effects and Operations: Fundamental Notions of Quantum Theory}, Lecture Notes in Physics, vol. 190, Springer-Verlag, 1983.
 
\bibitem{Mforsh} W. A. Majewski,  Fortschr. Phys., 32 (1984)  89 
  
\bibitem{A11} R. Alicki, K. Lendi, {\it Quantum Dynamical Semigroups and Applications}, Lecture Notes in Physics, vol. 286, Springer-Verlag, 1987.

\bibitem{P15} P. Pechukas, Phys. Rev. Lett. 73 (1994) 1060.

\bibitem{A20} R. Alicki, Phys. Rev. Lett. 75 (1995) 3020.

\bibitem{B?} F. Benatti, R. Floreanini, M. Piani, Open Syst. Inf. Dyn.11, (2004) 325

\bibitem{AF} R. Alicki, M. Fannes, {\it Quantum dynamical systems}, Oxford University Press, 2001

\bibitem{S17} T. F. Jordan, A. Shaji, E. C. Sudarshan, Phys. Rev. A 70 (2004) 052110

\bibitem{SPLett} A. Shaji, E. C. Sudarshan, Phys. Lett. A 341 (2005) 48.
  
\bibitem{M22} W. A. Majewski, Positive maps, states, entanglement and all that; some old and new problems, e-print \texttt{quant-ph/0411043}.

\bibitem{Nak} S. Nakajima, Prog. Theor. Phys. 20 (1958) 948

\bibitem{Zwan} R. Zwanzig, J. Chem. Phys. 33 (1960) 1338; Phys. Rev. 124 (1961) 983

\bibitem{Mori} H. Mori, Prog. Theor. Phys. 33 (1965) 423

\bibitem{Grab} H. Grabert, {\it Projection operator technique in Nonequilibrium Statistical Mechanics}, Springer-Verlag, Berlin, (1982)

\bibitem{Rau} J. Rau and B. M{\"u}ller, Phys. Rep. 272 (1996) 1

\bibitem{Av} W. B. Arverson, Acta Math. 123 (1969) 141

\bibitem{EvaL} D. E. Evans and J. T. Lewis, {\it Dilations of Irreversible Evolutions in Algebraic Quantum Theory},
Dublin Institute for Advanced Studies, 1977

\bibitem{Ste} E. St{\o}rmer, J. Funct. Anal. 66 (1986) 235 

\bibitem{Dirac} P. A. M. Dirac, {\it The Principles of Quantum Mechanics}, Clarendom, Oxford, third edition 1947

\bibitem{JvN} J. von Neumann, {\it Mathematical Foundations of Quantum Mechanics}, Princeton University Press, 1955 

\bibitem{JNW} P. Jordan, J. von Neumann, E. Wigner, Ann. Math. 35 (1934) 29

\bibitem{Seg} I. E. Segal, Ann. Math., 48 (1947) 930

\bibitem{Mackey} G. Mackey, {\it Mathematical Foundations of Quantum Mechanics}, Benjamin, 1963

\bibitem{Emch} G. G. Emch, {\it Algebraic Methods in Statistical Mechanics and Quantum Field Theory}, John Wiley $\&$ Sons, Inc. 1972

\bibitem{Ludwig} G. Ludwig, {\it An Axiomatic Basis for Quantum Mechanics}, Springer-Verlag, 1985

\bibitem{CMP} L. J. Bunce, J. D. Wright, Commun. Math. Phys. 98 (1985) 187

\bibitem{St} E. St{\o}rmer, Math. Ann. 247 (1980) 21

\bibitem{BR} O. Bratteli, D.W. Robinson, {\em Operator Algebras and Quantum Statistical 
Mechanics}, Springer Verlag, Vol.I
(1979), Vol.II (1981)

\bibitem{JW} P. Jordan, E. Wigner, Z. Phys. 47 (1928) 631

\bibitem{Bon} P. J. Bongaards, Linear fields according to I. E. Segal, in {\it Mathematics of Contemporary Physics}, Ed. R. F. Streater, Academic Press, 1972

\bibitem{BSZ} J. C. Baez, I. E. Segal. Z. Zhou, {\it Introduction to Algebraic and Constructive Quantum Field Theory},
Princeton Series in Physics, 1992

\bibitem{Kit} S. B. Bravyi, A. Y. Kitaev, Ann. Phys. 298 (2002) 210

\bibitem{Vin} B. M. Terhal, D. P. DiVincenzo, Phys. Rev. A 65 (2002) 032325

\bibitem{HOS} H. Hanche-Olsen, E. St{\o}rmer, {\it Jordan Operator Algebras}, Pitman Advanced Publishing Program, 1984

\bibitem{ESt} E. G. Effros and E. St{\o}rmer, Math. Scand. 45 (1979) 127

\bibitem{Ar1} H. Araki, H. Moriya, Commun. Math. Phys. 237 (2003) 105

\bibitem{Ar2} H. Moriya, Lett. Math. Phys. 60 (2002) 109

\bibitem{Ume} H. Umegaki,  Tohoku Math. J. 6 (1954) 177

\bibitem{Tak} M. Takesaki, {\it Theory of Operator Algebras I}, Springer-Verlag, 1979

\bibitem{AC} L. Accardi and C. Cecchini, 
 J. Func. Anal., 45 (1982) 245

\bibitem{Wein} Y. S. Weinstein, T. F. Havel, J. Emerson, N. Boulant, M. Saraceno, S. Lloyd and D. G. Cory, J. Chem. Phys. 121 (2004) 6117

\bibitem{Moriya} H. Moriya, J. Phys. A: Math. Gen. {\bf 39} (2006) 753
\end{thebibliography}
\end{document}